\documentclass[11pt]{emulateapj}\usepackage{times}
\begin{document}

\shorttitle{\textsc{GLAST testing of a pulsar model of LS 5039 }}
\shortauthors{\textsc{A. Sierpowska-Bartosik \& D. F. Torres}}

\title{\textsc{%The nature of the binary LS 5039: 
GLAST testing of a pulsar model matching H.E.S.S. observations of LS 5039 }}
%\received{} \revised{}    \accepted{}
\author{Agnieszka Sierpowska-Bartosik\altaffilmark{1} \& Diego F. Torres\altaffilmark{2,1}}

\altaffiltext{1}{Institut de Ci\`encies de l'Espai (IEEC-CSIC),
              Campus UAB,  Torre C5, 2a planta,
              08193 Barcelona, Spain.
              Email: agni@ieec.uab.es}
\altaffiltext{2}{Instituci\'o Catalana de Recerca i Estudis Avan\c{c}ats (ICREA).
              Email: dtorres@ieec.uab.es}

\begin{abstract}
LS 5039 is one of a handful of X-ray binaries that have been recently detected at high-energy $\gamma$-rays, in this case, by the High-Energy Stereoscopy Array (H.E.S.S.). The nature of this system is unknown: both a black hole and a pulsar have been invoked as possible compact object companions. Here we work with a model of the high energy phenomenology of the system  in which it is assumed that the companion object is a pulsar rotating around an O6.5V star in the $\sim 3.9$ days orbit. The model assumes two different sets of power-law spectral parameters of the interacting primary leptons corresponding to the two orbital phase intervals defined by H.E.S.S. as having different gamma-ray spectra and very-high-energy (VHE) cutoffs.  We show the H.E.S.S. phenomenology is completely explained by this model. We present predictions for photons with lower energies (for $E>1 $ GeV), subject to test in the forthcoming months with the GLAST satellite. We find that GLAST is able to judge on this model within one year. 
\end{abstract}


\section{Introduction}

The Gamma-ray Large Area Space Telescope (GLAST) is the next generation  $\gamma$-ray observatory due for launch in May 2008.
Its primary instrument is the Large Area Telescope 
(LAT), which will measure $\gamma$-ray flux and spectra from 100 MeV to $\sim$100 GeV. LAT is the 
successor of the EGRET experiment that flew onboard the Compton satellite.
LAT will have better angular 
resolution, greater effective area, wider field of view, and broader energy coverage than EGRET. 
Developing predictions for the GLAST energy domain upon which to test models of LS 5039 is then timely and worth exploring. 

In the previous work (Sierpowska-Bartosik \& Torres 2007, where we refer the reader for  references), under the assumption that LS 5039 is composed by a pulsar rotating around an O6.5V star in the $\sim 3.9$ days orbit,  
we presented the results of a theoretical modeling of the high energy phenomenology observed by  H.E.S.S.\footnote{H.E.S.S. found a periodicity in the $\gamma$-ray flux --consistent with the orbital timescale as determined by Casares et al. (2005)-- and fast variability displayed on top of this periodic behavior, both in flux 
and spectrum (Aharonian et al. 2006).}
The main difference between the previous model and the current one is that the former assumed a constant spectrum describing the interacting particles along the orbit, whereas now it is let to vary.
The previous model (including detailed account of the system geometry, Klein-Nishina inverse Compton (IC), $\gamma \gamma$ absorption, and cascading)  was able to describe reasonably well the rich details 
found in the system, both flux and spectrum-wise. However, almost the complete spectral energy distribution at superior conjunction, as well as other features of  the observed orbital variation of the H.E.S.S. energy spectra,  remained to be consistently explained. With models that do not entirely match the observed data at the highest energies, the study of lower energy predictions lacks testing power (strictly speaking, these models are already ruled out by H.E.S.S. observations). Here, we analyze and motivate the description of the VHE data allowing for a variable spectrum of interacting primary leptons along the orbit and indicate that the new model can nicely explain the H.E.S.S. observations.  We then present predictions for the flux and spectrum as function of phase for photons with lower energies, where H.E.S.S. is not sensitive but GLAST is.
LS 5039 
is near the galactic center and has two other $\gamma$-ray sources nearby, as well as a significant galactic diffuse 
background: it is not an easy target for GLAST (see Dubois 2006).  
In order to study orbital variability, the
need to integrate long time intervals and 
make harder energy cuts to bring the signal above background was reported.

\section{The pulsar model with 
orbital variability in the interacting primary spectrum }

The main features of the model are described by Sierpowska-Bartosik \& Torres (2007). For LS 5039's parameters the star wind dominates over the putative pulsar's, and a shock wraps around the latter.  We assume then that the volume of the system is shock-separated as a result of the collision between the pulsar and the massive star winds. 
Between the pulsar and the shock, inside the pulsar wind zone (PWZ), relativistic leptons are assumed to be frozen in the magnetized pulsar wind which propagates  radially from the pulsar. 
In this PWZ, we compute Klein-Nishina IC cascades against thermal radiation from the massive star. Secondary $\gamma$-rays move into the massive star wind region and whereas some escape the binary system, others get absorbed due to $\gamma\gamma$ process. These cascades are followed by means of a Monte Carlo procedure. We have computed geometric dependences upon the opacities to these processes. For close binaries, the radiation field of hot massive stars (type O, Be or WR, having typical surface temperatures in the range $T_s \sim 10^4 - 10^5\, \rm K$ and linear dimension $R_s \sim 10 R_{\odot}$) dominates along the whole orbit over other possible fields (e.g., the magnetic field or the thermal field of the neutron star). This thermal radiation field is anisotropic for $e^+e^-$ injected close to the pulsar (the radiation source is misplaced with respect to the electron injection place).

We assume that the injected power in relativistic leptons, which initiate the cascading processes, is a fraction of the spin-down luminosity, $L_{\rm SD}$. We particularize on the simplest case in which leptons are described, after being reprocessed by losses which dominance can be a function of orbital phase, by a power-law in energy that may be constant (as in Sierpowska-Bartosik \& Torres 2007) or vary along the orbit. Here, we focus on  the case in which two different power-laws are assumed for the interacting lepton  population, corresponding to the two orbital intervals proposed by H.E.S.S. around inferior  (0.45 $< \phi\leq$ 0.90) and superior ($\phi\leq 0.45$ and $\phi>0.90$) conjunction (INFC and SUPC). Table 1 shows the model parameters chosen. In both cases we assume a nominal value for $L_{SD}=10 ^{37}$ erg s$^{-1}$, and $E_{\rm max} $=50 TeV.
Note that  $L_{\rm SD}$ and the fraction of it that is converted into relativistic leptons ($\beta$) are obviously related.  The position of the shock (which defines the size of the PWZ where we compute the cascades) is defined by the parameter $\eta =  L_{\rm SD} /(\dot{M} V_{\rm w}c)$, where the wind velocity, $V_{\rm w}$, depends on the radial distance from the 
massive star, and $\dot{M}$ is the star mass-loss rate. Then,  $\dot{M}V_{\rm w}$ and $L_{\rm SD} $ are also connected. Different sets of parameters (e.g., a smaller $L_{\rm SD}$ with a higher $\beta$) give similar results. This is the consequence of a mild dependence with $\eta$ of the distance from the pulsar to the shock. 
In the models presented here, only a small fraction
($\sim $1\%) 
of the pulsar's $L_{\rm SD}$ ends up in relativistic leptons. 
This is consistent with ions carrying much of the wind luminosity.

Once the pulsar injects relativistic leptons (what could itself  
be subject to orbital variability), or alternatively,
a close-to-the-pulsar shock accelerates a primary population 
assumed to be isotropic in the pulsar rest frame (e.g., Kirk et al. 1999),
the interacting lepton population arises from 
the equilibrium between injection and the losses operating in the system along the orbital evolution. 
In this sense, SUPC and INFC of LS 5039 may indeed represent qualitatively different physical scenarios, as noted already by Aharonian et al. (2006). 
Let's consider first the maximum energy to which leptons are accelerated in the case the latter proceeds in a shock. 
The high temperature of the star and the hard spectrum of the VHE radiation measured imply (unless a very hard injection proceeds) that IC must happen in the Klein-Nishina regime. For dominant IC cooling, shock acceleration and cooling timescales equality implies that 
$E_{\rm max}  \propto (B/w)^{3.3}$, where $B$ is the magnetic field and $w$ is the target photon energy density. Since $w \propto d^{-2}$, with  $d$ the binary separation, and  $B$ is a decreasing function of $d$, e.g., $B \propto d^{-1}$, 
$E_{\rm max} $ increases by a factor $\sim 10$ from 
periastron (in the SUPC broad-orbital range) to apastron (in the INFC broad-orbital range), leading to a spectral hardening around INFC. 
At the highest energies, synchrotron losses (which timescale is $\propto E^{-1}$) are expected to dominate and steepen the spectrum. The changeover energy from IC to synchrotron domination of the radiative losses depend on $B$ and $d$; if $B$ is not exactly $\propto 1/d$, it may also introduce a spectral hardening at apastron.
Finally, adiabatic losses happening with the typical scales of the system
associated with the flow patterns --i.e., 
the stand-off distance of the pulsar wind shock (which scales with, and is less than the 
distance between
the stars), divided by the
post-shock speed ($\sim c/3$)-- generates a timescale that is independent of energy
and, for LS 5039, close to or even smaller than the Klein-Nishina cooling times.
In general, if leptons enter the production region with a rate described by a power-law with index $\alpha_0$, in the case of dominant radiative losses either by synchrotron or Thompson IC, the slope of the interacting population is steppened to $\alpha_0+1$. In the case of dominance by Klein-Nishina IC,
the decrease in the cross section is such that the 
distribution index is slightly modified or even hardened at most by  
$< 0.5$ in index, with hardening decreasing with the ratio $U_\star/U_B$ (Moderski et al. 2005). In the case of adiabatic dominance, 
 the slope of the interacting population 
 remains the same. Given that the losses dominance along LS 5039's orbit can be a function of phase, and in addition  that also the opacity to pair production is phase dependent, it seems natural to assume that in approximating the  distribution of the interacting
lepton population with a power-law, it has a different index along the two broad-phase intervals which qualitative observed features differ the most. 
%The predictions of such a model are what we test here. 
%
In this context, the distinction between electron spectra at INFC and SUPC should  be understood as an average of a  smoother phase-dependence of the electron primary spectrum, what could in turn be tested with future quality of data if more complex modeling is available.

\section{Matching of H.E.S.S. results and predictions for  
GLAST}

Fig.  1 shows
the results of our model, both flux and spectrum-wise, compared with corresponding H.E.S.S. data  (obtained from Aharonian et al. 2006). 
The spectra for LS 5039 is presented in two broad orbital phase intervals around INFC and SUPC.  Each of the experimental data points represents a typical time span of 28 min. The corresponding phases for apastron, periastron, INFC, and SUPC are marked. Two periods are shown. To ease the comparison with our earlier result using a constant spectrum of interacting particles along the orbit (Sierpowska-Bartosik \& Torres 2007) we include it  (with a finer gridding) also in these plots. 
H.E.S.S. has also provided the evolution of the normalization and slope of a power-law fit to the 0.2--5 TeV data in 0.1 phase-binning along the orbit (Aharonian et al. 2006). The use of a power-law fit was limited by low statistics in such shorter sub-orbital intervals, i.e., higher-order functional fittings such as a power-law with exponential cutoff were reported to provide a no better fit and were not justified. To directly compare with these results, we have applied the same approach to treat the model predictions, i.e., we fit a power-law in the same energy range and phase binning. This comparison is shown in Fig.  \ref{phase-bin}. We find a rather good agreement between model predictions and data.

Fig.  1 also shows the spectral energy distribution predictions extended for energies above 1 GeV. The steeper the primary spectrum is, the higher the flux produced at lower energies, what is particularly notable for the SUPC broad phase-interval (see bold and thin lines of Fig.  1).
The minimum flux
needed for a source to be detected by GLAST after a 1-month
and 1-year of operation in all-sky survey, for a $E^{-2}$ source.\footnote{www-glast.slac.stanford.edu/software/IS/glast\_lat\_performance.htm}
At these flux levels,  a 20\%-uncertainty in the determination of the flux, a resulting significance  about 8$\sigma$, and a spectral index determined to about 6\% would be achieved. Even when these sensitivities maybe slightly worse for a low-latitude source,  if this model is correct,  GLAST should be able to quantify the orbital variability after a few months of integration. The fact that the system periodicity is $\sim$3.9 days allows for a fast build-up of integration time around each of the portions of the orbit. A month-integration around SUPC will be obtained after $\sim$2 months of all-sky survey, putting this model to the test soon after GLAST launch.\footnote{LS 5039 is positionally coincident with 3EG J1824-1514 (Hartman et al. 1999, Paredes et al. 2000), whose average $\gamma$-ray flux above 100 MeV 
%--along all EGRET viewing periods-- 
is $\sim 3.5 \times 10^{-7}$ photons cm$^{-2}$ s$^{-1}$. Our model predictions 
%for this energy range 
are consistent with this observation too.}
Our model also predicts a clear anti-correlation between the expected output at TeV and GeV energies, which can be seen comparing the lightcurves shown in Figs.  1 and 3. The lower the energy range, the more anti-correlated the signal is, what is a consequence of the phase-dependence of the cascading and absorption processes. A hardness ratio defined within the GLAST energy domain does not vary as much as it does when constructed combining the fluxes at low and high $\gamma$-ray energies (a factor of ~3 versus an order of magnitude), but it may be useful as a first check before integration time is achieved for determining the spectrum. This is shown in Fig. 4.

\section{Concluding Remarks}

A detailed modeling of LS 5039, under the assumption that it contains a pulsar, is able to fully describe the challenging H.E.S.S.-observed phenomenology found in the system. Observed lightcurve, spectra (Fig.  1), and short-timescale spectral variability (Fig.  2) are matched with a variable spectrum of primary particles along the orbit, where two phase intervals are considered. It is interesting to see that the influence of the orbital inclination angle is minor along all energy intervals within our theoretical model. Predictions of the model at lower $\gamma$-ray energy domains are ready for testing with GLAST. We find that GLAST is able to judge on this model within one year. Features both at the lightcurve and spectral level can be recognized. Additional tests will come at higher energy yet beyond the current reach, e.g., behavior of the spectra at shorter phase intervals to be compared with data from the next generation of ground-based instruments: H.E.S.S. II, and the planned Cerenkov Telescope Array (CTA).

%%%%%%%%%

%%%%%%%%
\acknowledgements
We acknowledge  the use of IEEC-CSIC computer cluster, J. M. Paredes and O. Reimer for comments, and support by grants
MEC-AYA 2006-00530 and CSIC-PIE 200750I029.

\clearpage

%table
\begin{deluxetable}{ll}
\tablewidth{.5\textwidth}
\tabletypesize{\small}
\tablecaption{Model Parameters}
\tablehead{\colhead{Parameter} & \colhead{Adopted value}}  
\startdata
{\bf  Constant lepton spectrum along the orbit} & \\
\hline
 Fraction of $L_{\rm SD}$ in leptons: $\beta$ & $10^{-2}$ \\  
 Slope of the power-law: $\Gamma_e$ & $-2.0$ \\
 \hline
{\bf Variable lepton spectrum along the orbit}&\\
\hline
 Fraction of $L_{\rm SD}$ in leptons at INFC  interval: $\beta$ &  $8.0 \times 10^{-3}$ \\  
 Slope of the power-law at INFC  interval: $\Gamma_e$ & $-1.9$ \\
  Fraction of $L_{\rm SD}$ in leptons at SUPC  interval: $\beta$ & $2.4 \times 10^{-2}$ \\  
 Slope of the power-law  at SUPC  interval: $\Gamma_e$ & $-2.4$ \\
\vspace{-2mm}
\enddata
\end{deluxetable}

\clearpage

\begin{figure*}
\centering
\includegraphics[width=.49\textwidth]{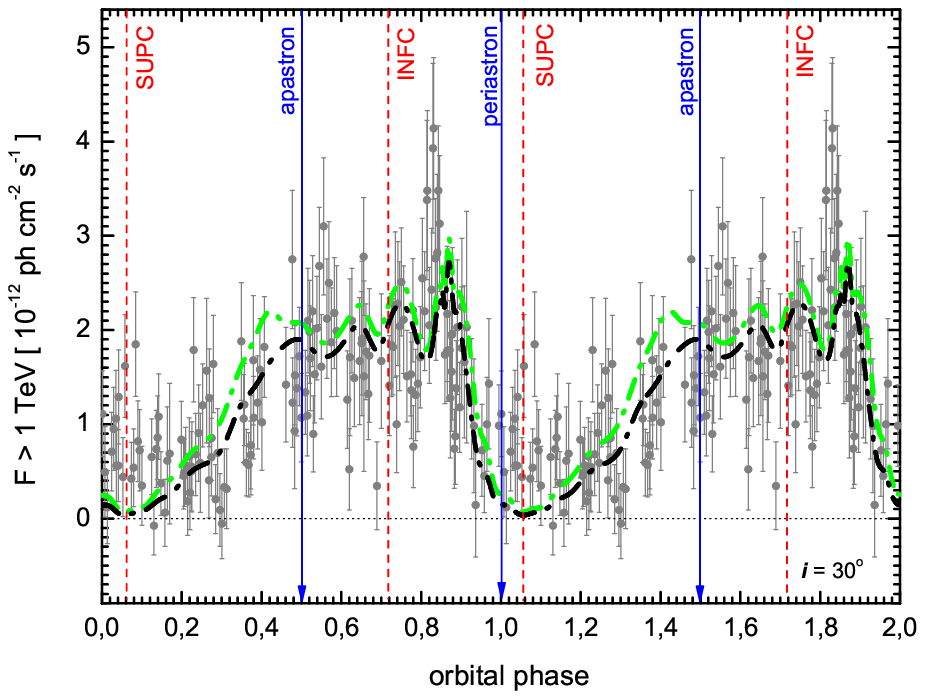}
\includegraphics[width=.49\textwidth]{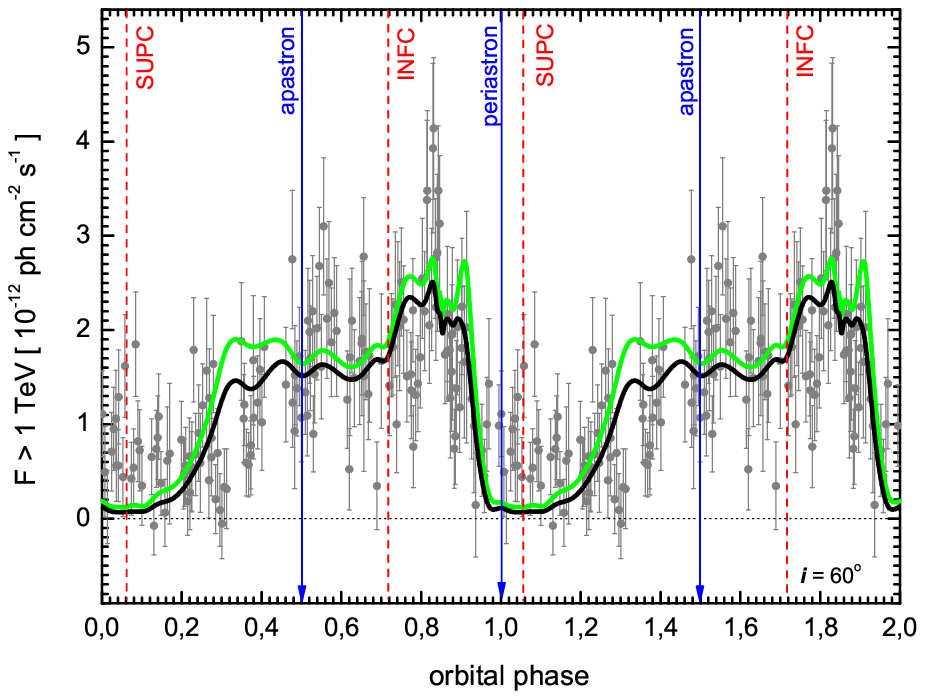}\\
\includegraphics[width=.49\textwidth,trim=0 5 0 10]{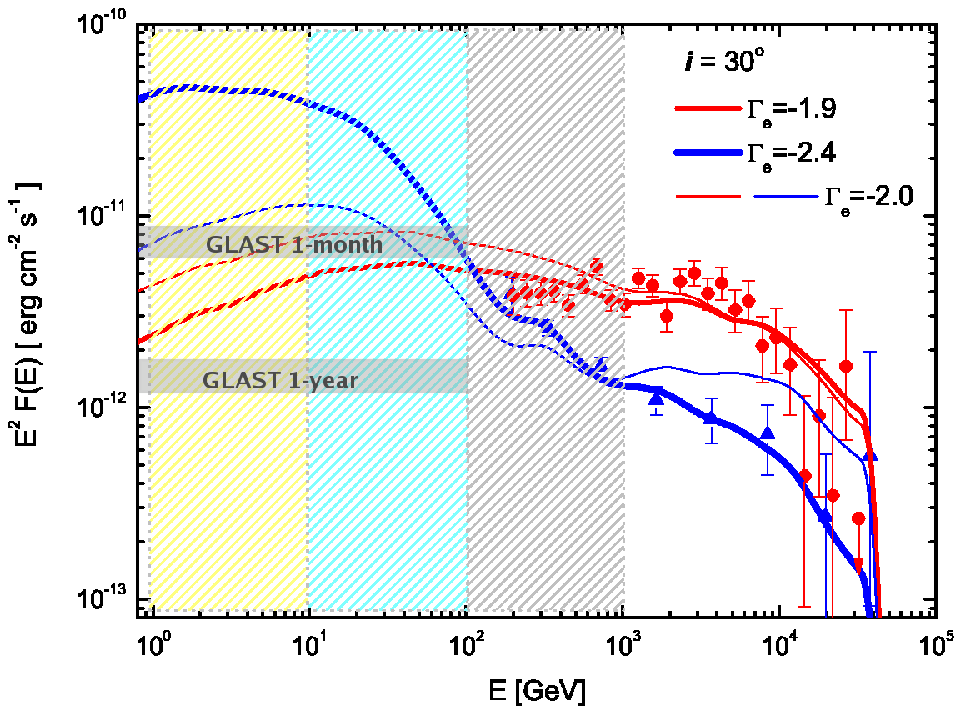}
\includegraphics[width=.49\textwidth, trim=0 5 0 10]{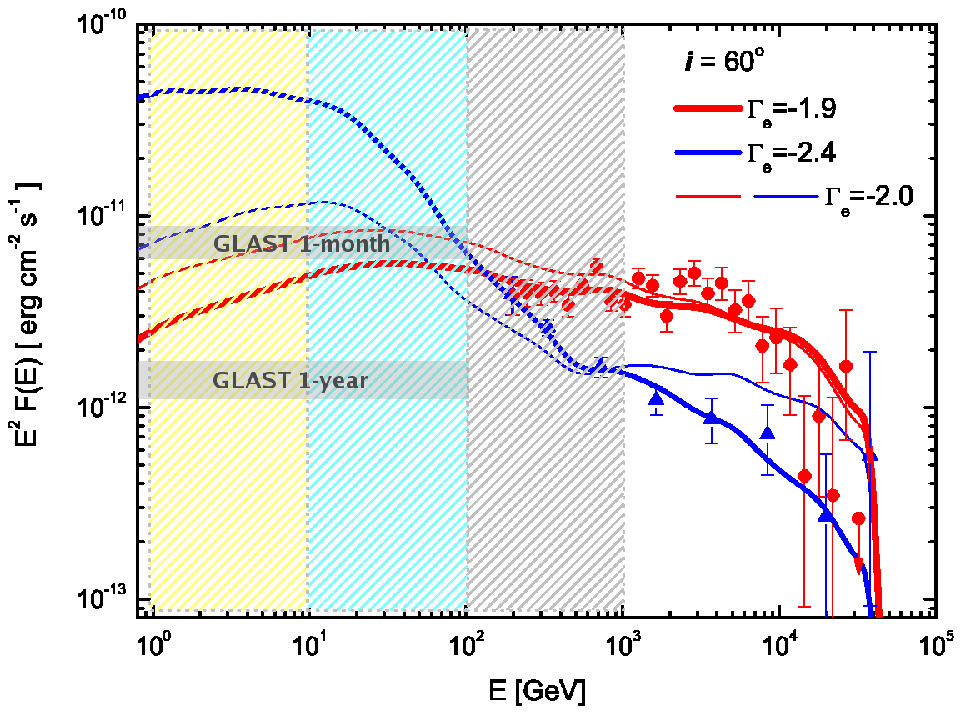}
\caption{Top: Lightcurves for photons with energy above 1 TeV for different binary inclinations (dot-dashed: $i=30^0$, solid lines: $i=60^0$), compared with  H.E.S.S. data. Black lines stand for results obtained with a variable interacting spectrum along the orbit; green (light) lines correspond to the constant spectrum case. Bottom: Spectra around INFC (in red) and SUPC (in blue). H.E.S.S. data is shown in the same colors. Results for both cases of interacting electron spectra are given. Shaded regions are energy ranges for which we study the lightcurves below. The two horizontal lines between 1 and 100 GeV represent sensitivities of GLAST in the all-sky survey mode. }
\label{LC}
\end{figure*}
\clearpage

\begin{figure*}
\centering
\includegraphics[width=.49\textwidth]{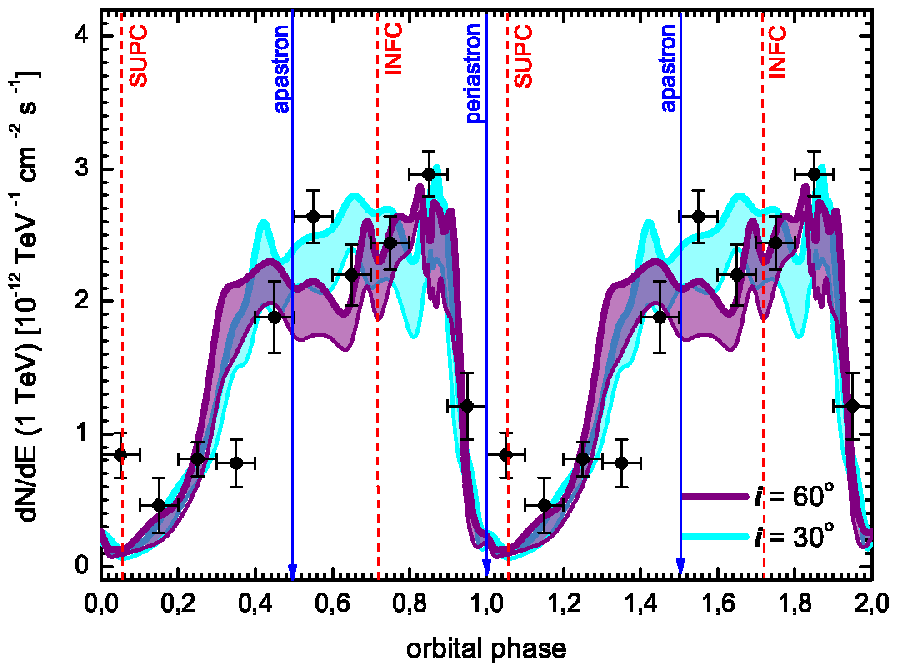}
\includegraphics[width=.49\textwidth]{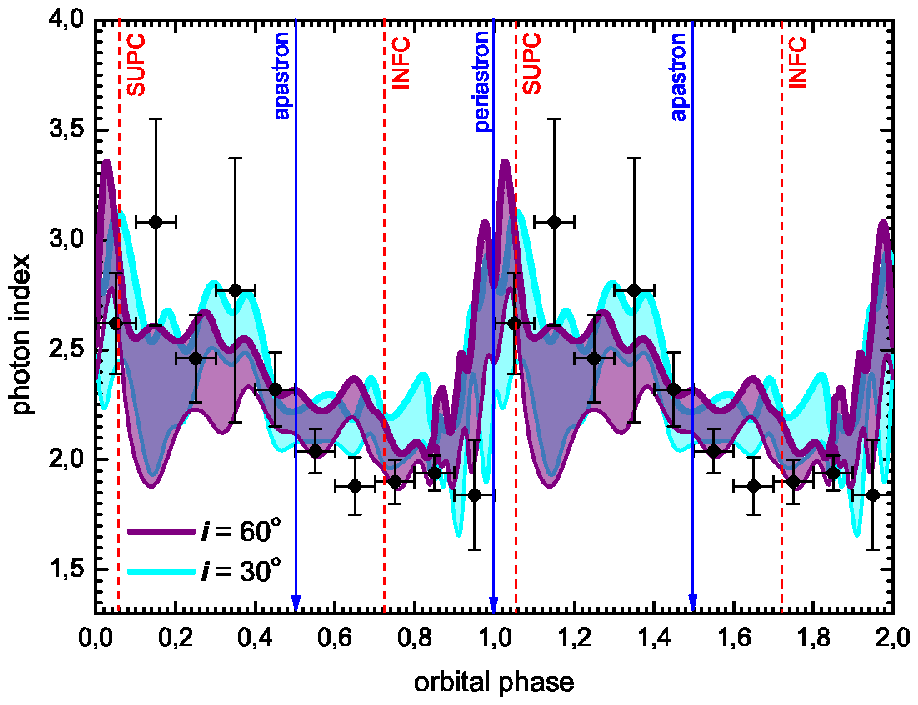}
\caption{Shaded areas in the left (right) panel show the change in the normalization (photon index) of a power-law photon spectra fitted to the theoretical prediction for each of the 0.1 bins of phase. The presented results are for the model of the variable interacting spectrum. The two different colors of the shading stand for the two inclination angles considered. The size of the shading gives account of the error in the fitting parameters. Data points represent the H.E.S.S. results for the equal procedure: a power-law fit to the observational spectra obtained in the same phase binning.}
\label{phase-bin}
\end{figure*}

\clearpage

\begin{figure*}
\centering
\includegraphics[width=.32\textwidth]{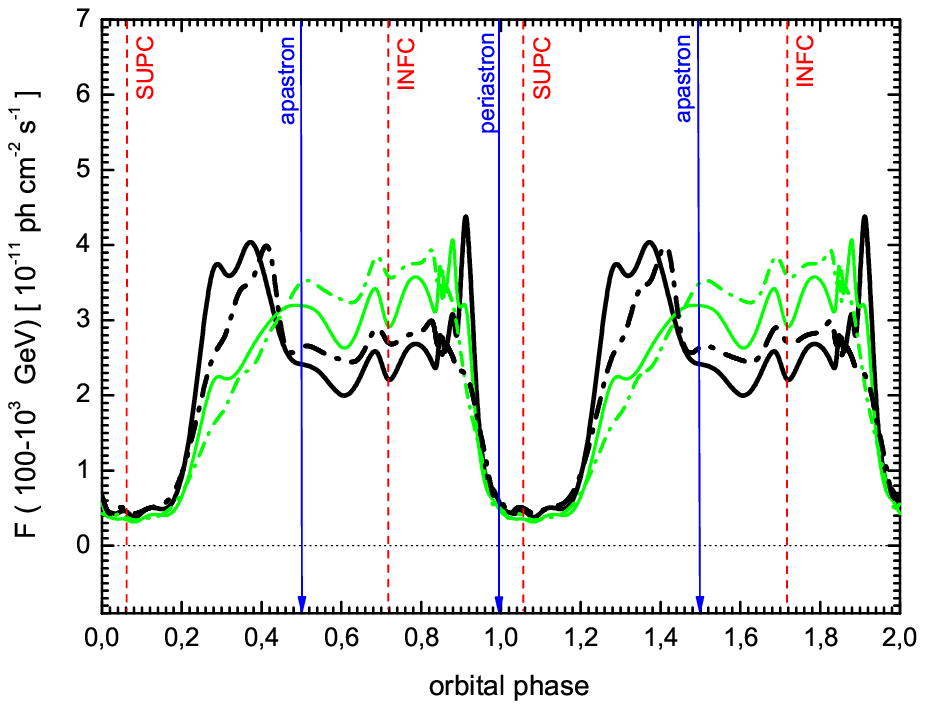}
\includegraphics[width=.32\textwidth]{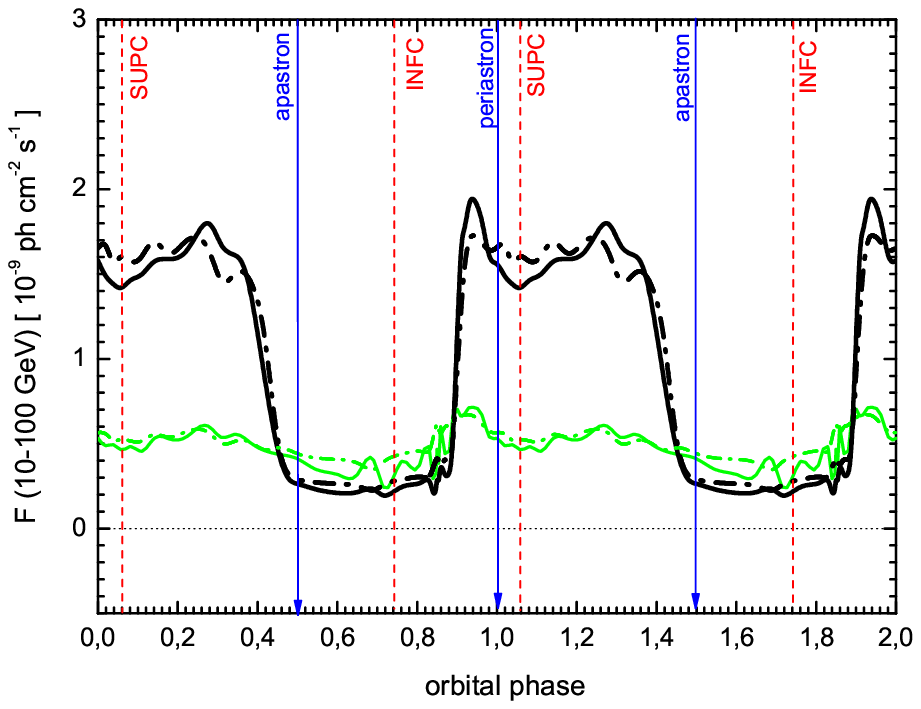}
\includegraphics[width=.32\textwidth]{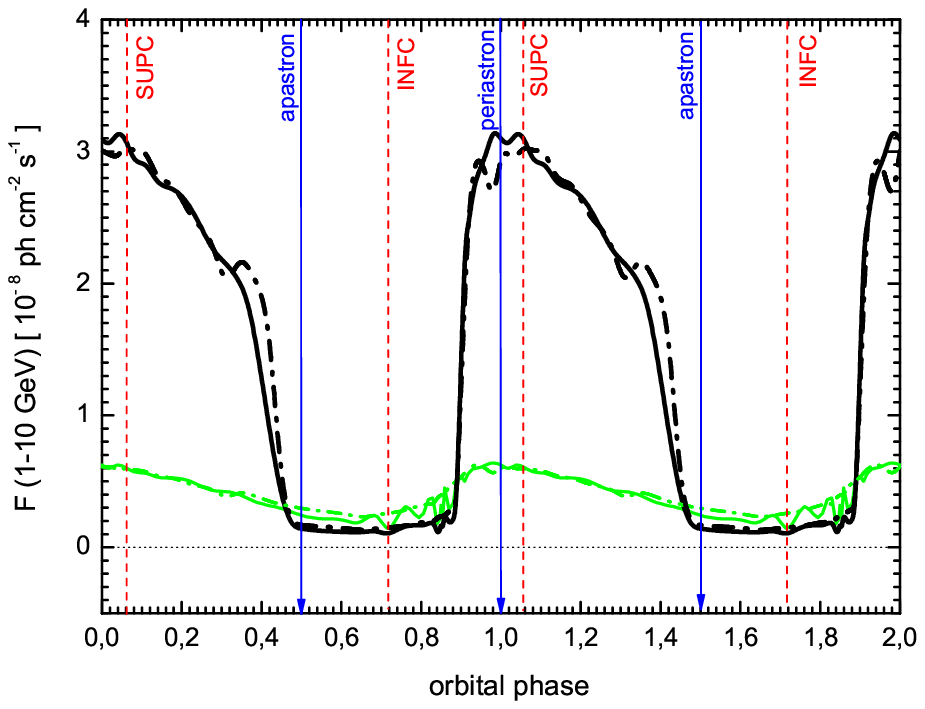}
\caption{Predicted theoretical lightcurves for the energy intervals 100 GeV -- 1 TeV, 10 -- 100 GeV, and 1--10 GeV. Both inclination angles and interacting lepton spectra considered are shown  (dot-dashed: $i=30^0$, solid lines: $i=60^0$, black lines: variable spectrum of primary leptons, green (light) lines: constant spectrum along the orbit). 
}
\end{figure*}

\clearpage

\begin{figure*}
\centering
\includegraphics[width=.49\textwidth]{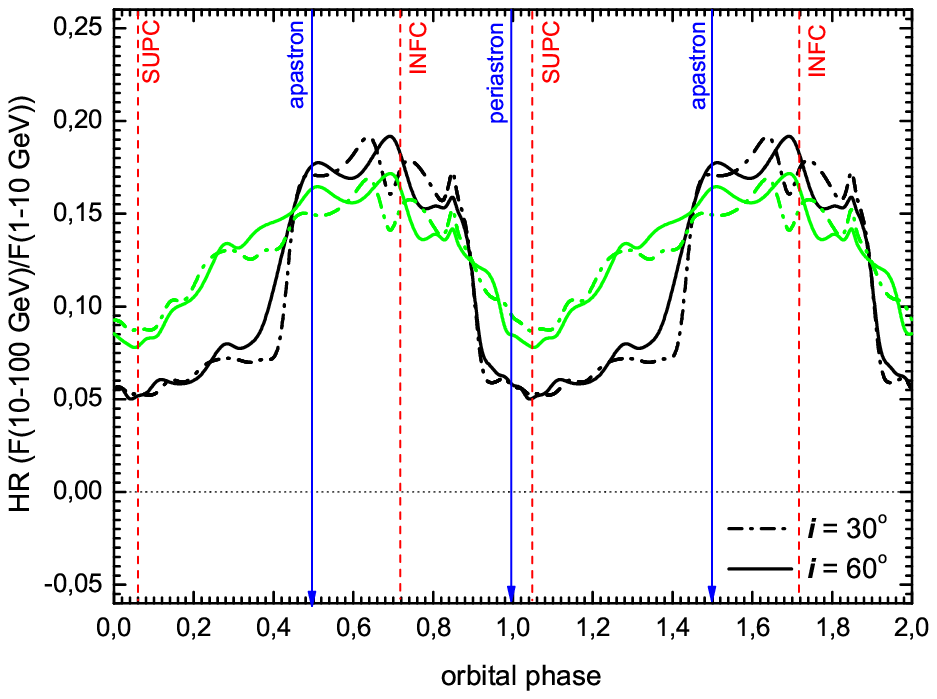}
\includegraphics[width=.49\textwidth]{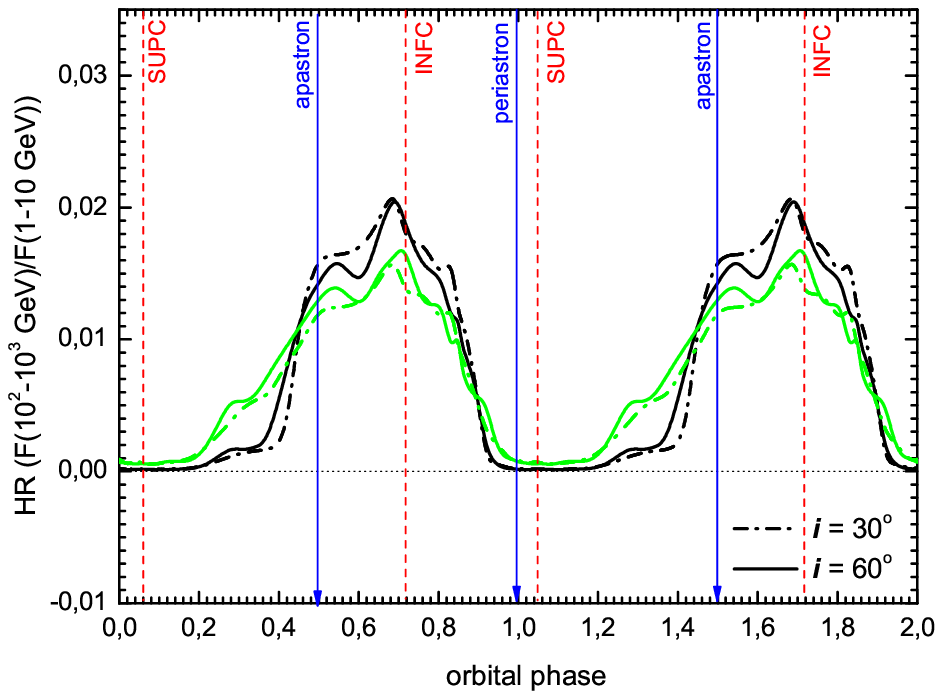}
\caption{Hardness ratios as a function of orbital phase, for two inclination angles (dot-dashed: $i=30^0$, solid lines: $i=60^0$). 
Color style follows that used in previous figures. The energy regimes  considered are $F({\rm 1 - 10 \, GeV})$, $F({\rm 10 - 100\, GeV})$, and $F({\rm 10^2 - 10^3 GeV})$. 
}
\label{HR}
\end{figure*}

\end{document}